\documentclass[prl,final,twocolumn,showpacs,amsmath,superscriptaddress]{revtex4}
\pdfoutput=1
\usepackage{hyperref}
\usepackage{graphicx}

\DeclareGraphicsExtensions{.pdf}

\newcommand{\figwidth}{\linewidth}
\newcommand{\br}{\mathbf{r}}
\newcommand{\dif}{\mathrm{d}}
\newcommand{\SC}{\mathcal{S}}
\newcommand{\etal}{\textit{et al.}}

\begin{document}

\title{What is measured in the scanning gate microscopy of a quantum point
contact?}

\author{Rodolfo A.\ Jalabert}
\affiliation{Institut de Physique et Chimie des Mat{\'e}riaux de 
Strasbourg, UMR 7504, CNRS-UdS,\\
23 rue du Loess, BP 43, 67034 Strasbourg Cedex 2, France}
\author{Wojciech Szewc}
\affiliation{Institut de Physique et Chimie des Mat{\'e}riaux de 
Strasbourg, UMR 7504, CNRS-UdS,\\
23 rue du Loess, BP 43, 67034 Strasbourg Cedex 2, France}
\author{Steven Tomsovic}
\altaffiliation[Permanent address:]{Department of Physics and Astronomy, P.O.\ Box
642814, Washington State University, Pullman, WA 99164-2814, USA}
\affiliation{Department of Physics, Indian Institute of Technology Madras,
Chennai, 600 036 India}
\author{Dietmar Weinmann}
\affiliation{Institut de Physique et Chimie des Mat{\'e}riaux de 
Strasbourg, UMR 7504, CNRS-UdS,\\
23 rue du Loess, BP 43, 67034 Strasbourg Cedex 2, France}

\date{\today}

\begin{abstract}
The conductance change due to a local perturbation in a phase-coherent
nanostructure is calculated. The general expressions to first and second order
in the perturbation are applied to the scanning gate microscopy of a
two-dimensional electron gas containing a quantum point contact.  The
first-order correction depends on two scattering states with electrons incoming
from opposite leads and is suppressed on a conductance plateau; it is
significant in the step regions.  On the plateaus, the dominant second-order
term likewise depends on scattering states incoming from both sides.  It is
always negative, exhibits fringes, and has a spatial decay consistent with
experiments. 
\end{abstract}

\pacs{85.35.Ds, 
     07.79.-v, 
     73.23.-b, 
     72.10.-d  
}

\maketitle


Scanning gate microscopy (SGM) has been intensively used during the past decade
to investigate electronic transport in nanostructured two-dimensional electron
gases such as quantum point
contacts~\cite{topinka00a,topinka01a,leroy05a,jura09a}, quantum
billiards~\cite{crook03b}, and rings~\cite{martins07a,pala08a}.  In this
technique, a charged tip that locally influences a device's electrons is scanned
over the surface; i.e. the perturbed nanostructure's conductance is a function
of the tip position.  The SGM technique is of great fundamental interest since
it provides a wealth of data that depends on the microscopic electron transport.
It can generate very detailed sample characterization and more precise
information of the disorder configuration than a traditional transport
measurement.

Of importance in SGM studies is the precise physical interpretation of data.  
Are the measurements sensitive to the electron flow as often just
stated or, perhaps, the local electron density?  Whereas experiments and
numerical simulations have yielded conductance change patterns caused by
local perturbations closely related to calculated electron flows or local
electron
densities~\cite{topinka01a,leroy05a,martins07a,pala08a,metalidis05,cresti06,he02}, 
the precise relationships among these quantities and interpretations remain far 
from obvious.  Moreover, it is recognized that in a linear response framework the 
current density is not uniquely defined, and its study merely provides a way to 
visualize the structure of the scattering states at the Fermi 
energy~\cite{baranger91a}. Thus, our goal is to provide a clear-cut relationship 
between the conductance change in a SGM setup and the unperturbed scattering states 
which is valid for an arbitrary coherent structure.  

Consider a system with a Hamiltonian $H=H_0+V$, where $H_0$ represents the 
unperturbed structure and $V$ the tip potential.  Though quite general, the main 
focus is on a quantum point contact (QPC). This paradigm exhibiting conductance 
quantization~\cite{vanwees88,wharam88,glazman,szafer89,buttiker90} has been well 
studied by SGM~\cite{topinka00a,topinka01a,leroy05a,jura09a}.  In addition, recent 
work suggests the relevance of this powerful technique for probing the nonlocality 
of electronic interactions~\cite{ferry08a,freyn08a,weinmann08a}.

The scattering theory of quantum conductance~\cite{jalabert00,mello04} assumes
that the leads are disorder-free, confined in the transverse ($y$) direction,
and semi-infinite in the longitudinal ($x$) direction. The electron states in
the leads are products of quantized transverse wavefunctions $\phi_{a}(y)$
labeled by the index $a$ with transverse energy $\varepsilon^{(\mathrm{t})}_a$ 
and plane waves propagating in the $x$ direction with wave-vector magnitudes $k_a$.  
The total electron energies are 
$\varepsilon=\varepsilon^{(\mathrm{t})}_a + \hbar^2 k^{2}_{a}/2M$, with $M$ the 
effective electron mass. The $N$ transverse momenta which satisfy this relationship 
with $k_{a}^2 > 0$ define the $2N$ propagating channels of the leads with energy 
$\varepsilon$.  Denoting $\br=(x,y)$, the incoming (-) and outgoing (+) lead states 
are defined as 
\begin{equation}
\varphi_{1,\varepsilon,a}^{(\pm)}(\br)  =  
\frac{\exp(\mp i k_{a}^\pm x)}{\sqrt{2\pi\hbar^2k_{a}/M}} 
\ \phi_{a}(y) \ , \quad x <  \ 0 \ .
\end{equation}
For $x>0$, the index $1\rightarrow 2$ and the opposite sign is taken in
the argument of the exponential.  An infinitesimal negative (positive)
imaginary part is given to $k_a$ for incoming (outgoing) lead states. For
simplicity, consider a confining potential that is $x$-independent in the
asymptotic region even though any separable potential can be treated if
allowance is made for an $x$ dependence of $k_a$.  

The scattering states corresponding to an electron incoming from the left lead
1 (right lead 2) with energy $\varepsilon$ in the mode $a$ are given in the
asymptotic regions by
\begin{eqnarray}
\Psi_{1,\varepsilon,a}^{(+)}(\br) &=& \left\lbrace \begin{array}{ll}
\varphi_{1,\varepsilon,a}^{(-)}(\br) + \sum_{b=1}^{N} r_{ba} 
\varphi_{1,\varepsilon,b}^{(+)}(\br), & x < 0 \nonumber \\
\sum_{b=1}^{N} t_{ba} \varphi_{2,\varepsilon,b}^{(+)}(\br) 
, & x > 0 \end{array} \right. \nonumber \\
\Psi_{2,\varepsilon,a}^{(+)}(\br) &=& \left\lbrace \begin{array}{ll}
\sum_{b=1}^{N} t^{\prime}_{ba} \varphi_{1,\varepsilon,b}^{(+)}(\br)
, & x < 0  \\
\varphi_{2,\varepsilon,a}^{(-)}(\br) + \sum_{b=1}^{N} r^{\prime}_{ba} 
\varphi_{2,\varepsilon,b}^{(+)}(\br), & x > 0 \end{array} \right. 
\end{eqnarray}
Then the $2N \! \times \! 2N$ scattering matrix $S$, relating incoming and
outgoing fluxes, can be written in terms of the $N \times N$ reflection and
transmission matrices from lead $1\ (2)$ as
\begin{equation}
S = \left( \begin{array}{cc}
r & t' \\
t & r'
\end{array} \right) \, .
\end{equation}
The chosen lead state normalization corresponds to an incoming flux $e/h$ and
ensures the orthonormality of the scattering states~\cite{jalabert00,mello04}. 
The zero-temperature conductance of the unperturbed structure in units of the
conductance quantum $2e^2/h$ is given by $g^{(0)}=\mathrm{Tr}[t^\dagger t]$,
where Tr denotes the trace over the modes.

The Lippmann-Schwinger equation for the perturbed wave function 
$\chi_{l,\varepsilon,a}^{(+)}(\br)$ using the retarded Green function 
$\mathcal{G}^{(0)}$ associated to $H_0$ is 
\begin{equation}\label{eq:Lippmann-Schwinger}
\chi_{l,\varepsilon,a}^{(+)}(\br)=\Psi_{l,\varepsilon,a}^{(+)}(\br)+
\int \dif \bar{\br} \ \mathcal{G}^{(0)}(\br,\bar{\br},\varepsilon)  
V(\bar{\br})  \chi_{l,\varepsilon,a}^{(+)}(\bar{\br}) \ .
\end{equation}
It is consistent within linear response theory to calculate the conductance change 
beginning with the current change carried by $\Psi_{1,\varepsilon,a}^{(+)}(\br)$ 
with the spectral decomposition of $\mathcal{G}^{(0)}$ in the scattering 
wave-function basis.  To first order in $V$, the integration over a cross section 
$\SC_x$ on the scatterer's right~\cite{fn-current-independent-of-x} gives 
\begin{widetext}
\begin{equation}
I_{1,\varepsilon,a}^{(1)}(x) = \frac{e\hbar}{M} \
\mathrm{Im}\left\lbrace\int_{\varepsilon^{(\mathrm{t})}_1}^{\infty} 
\frac{\dif\bar{\varepsilon}}{\varepsilon^{+}-\bar{\varepsilon}}
\sum_{\bar{l}=1}^{2}\sum_{\bar{a}=1}^{\bar{N}} 
Z^{1,\bar{l}}_{a,\bar{a}}(\varepsilon,\bar{\varepsilon}) \ 
\mathcal{V}^{\bar{l},1}_{\bar{a},a}(\bar{\varepsilon},\varepsilon)
\right\rbrace  \, ,
\label{eq:deltaI}
\end{equation}
where the matrix element of the perturbing potential in the scattering state basis is 
\begin{equation}
{\cal V}^{\bar{l},l}_{\bar{a},a}(\bar{\varepsilon},\varepsilon) =
\int \dif \br \
\Psi_{\bar{l},\bar{\varepsilon},\bar{a}}^{(+)*}(\br) \ V(\br) \ 
\Psi_{l,\varepsilon,a}^{(+)}(\br)\, . 
\end{equation}
In addition, a shorthand notation is introduced for the following quantities
involving the unperturbed states:
\begin{eqnarray}
Z^{1,1}_{a,\bar{a}}(\varepsilon,\bar{\varepsilon}) &=& \frac{iM}{2\pi\hbar^2}  
\sum_{b=1}^{\hat{N}}
\left(\sqrt{\frac{\bar{k}_b}{k_b}}+
\sqrt{\frac{k_b}{\bar{k}_b}}\right) 
t^{*}_{ba}t^{\phantom{*}}_{b\bar{a}}
\exp{\left[i(\bar{k}^{+}_b-k^-_b)x\right] } \, ,  \\
Z^{1,2}_{a,\bar{a}}(\varepsilon,\bar{\varepsilon}) &=& 
\frac{iM}{2\pi\hbar^2}
\left\lbrace
\left(\sqrt{\frac{k_{\bar{a}}}{\bar{k}_{\bar{a}}}}-
\sqrt{\frac{\bar{k}_{\bar{a}}}{k_{\bar{a}}}}\right)
t^{*}_{\bar{a}a}\exp{\left[-i(\bar{k}^{-}_{\bar{a}}+k^-_{\bar{a}}
)x\right ] }
+ \sum_{b=1}^{\hat{N}}
\left(\sqrt{\frac{\bar{k}_b}{k_b}}+
\sqrt{\frac{k_b}{\bar{k}_b}}\right)t^{*}_{ba}r'_{b\bar{a}}
\exp{\left[i(\bar{k}^{+}_b-k^-_b)x\right]}
\right\rbrace , \nonumber
\end{eqnarray}
\end{widetext}
with $\hat{N}= \mathrm{min}\lbrace N,\bar{N}\rbrace$.

The $\bar{\varepsilon}$ integral of Eq.~(\ref{eq:deltaI}) has a principal part
and a $\delta$-function contribution. Since the current correction has to be
independent of $x$, the integral can be evaluated in the limit $x\to\infty$. The
principal part contribution is simply equal to that of the $\delta$-function contribution.
The result is thus for $\bar{\varepsilon}=\varepsilon$, and forthwith the energy
arguments are dropped. The current change associated with the mode $a$ at energy
$\varepsilon$ is given by
\begin{equation}
I_{1,\varepsilon,a}^{(1)}  = \frac{e}{\hbar} \
\mathrm{Im}\left\lbrace \left(t^{\dagger}t\ \mathcal{V}^{1,1}
- r^{\dagger}t'\ \mathcal{V}^{2,1}\right)_{a,a}\right\rbrace  \, . 
\end{equation}
The first term vanishes when summed over $a$. Within
linear response in the bias voltage, the first-order change in the
zero-temperature conductance is
\begin{equation}
\label{eq:deltag1}
g^{(1)}  = - 4 \pi\ \mathrm{Im}\left\lbrace \mathrm{Tr}
\left[r^{\dagger}t'\ \mathcal{V}^{2,1}\right]\right\rbrace \, .  
\end{equation}
This result is valid for the general situation where quantum transport through a scatterer 
is modified by a weak perturbation~\cite{symmetry-of-dg}.  The matrix $r^\dagger t'$ 
depends only on the unperturbed scatterer, while the tip's effect appears in the 
$\mathcal{V}$ matrix elements. The conductance change is not simply given by the current 
density or charge density at the tip position but is proportional to products involving a 
matrix element of $V$ with two scattering states, one corresponding to an incoming electron 
from the left and the other from the right lead. The conductance's sensitivity to 
electrostatic potential variations has been considered in a one-dimensional geometry in 
Ref.~\cite{gasparian96}. The example of a $\delta$-function barrier perturbed by a local 
tip can be analytically calculated and agrees with Eq.~(\ref{eq:deltag1}). 

For a QPC, $r^\dagger t'$ is appreciable only in the vicinity of the conductance steps and 
is suppressed on the plateaus.  A basis transformation into the eigenmodes of $t^\dagger t$ 
singles out the linear combinations corresponding to the modes that propagate through the 
constriction~\cite{buttiker90}.  Only at these channels' openings is $r^\dagger t'$ 
significant. The first-order conductance correction [Eq.~(\ref{eq:deltag1})] is thus the 
relevant term in the step regions, but to capture the plateau regions' dominant 
correction~\cite{limit-low-tip-voltage}, the second-order contribution to the scattering 
wave functions in Eq.~(\ref{eq:Lippmann-Schwinger}) must be taken into account. Continuing 
with the same method gives the second-order conductance correction
\begin{eqnarray}
g^{(2)}  & = & - 4 \pi^2\   
\mathrm{Tr}\left[t^{\dagger}t\ \mathcal{V}^{1,2}\mathcal{V}^{2,1}
+ t'^{\dagger}t'\ \mathcal{V}^{2,1}\mathcal{V}^{1,2} \right. \nonumber \\
&&+\left.
\mathrm{Re}\left\lbrace r^{\dagger}t'\left(\mathcal{V}^{2,2}\mathcal{V}^{2,1}
-\mathcal{V}^{2,1}\mathcal{V}^{1,1}\right) \right\rbrace \right] \, . 
\label{eq:deltag2}
\end{eqnarray}
This result, together with Eq.~(\ref{eq:deltag1}), provides the answer to the question posed in the title.

Like $g^{(1)}$, $g^{(2)}$ depends on the scattering amplitudes of the QPC and the
scattering states incoming from both sides.  In contrast to $g^{(1)}$, $g^{(2)}$
contains nonvanishing terms with prefactors $t^{\dagger}t$ and
$t'^{\dagger}t'$. Since $r^{\dagger}t^\prime$ is suppressed on the plateaus,
$g^{(2)}$ is expected to be the dominant correction there. Furthermore, the
second-order terms contain contributions from all effective transmitting
channels, in contrast to $g^{(1)}$, which is significant only in the steps and
dominated by the opening mode. Thus, the transverse shape of the channels should
be observable directly with SGM on the steps without subtracting the lower mode results as done in the plateau data analysis~\cite{topinka00a}. 

Crucial insights into conductance quantization of a QPC have been gained
through the analysis of simple models, such as the saddle
potential~\cite{buttiker90} for the smooth electrostatic potential of the
constriction.  In addition, describing the saddle potential by a double harmonic
oscillator $V(\mathbf{r})=V_0+\frac{M}{2}(\omega^2_y y^2-\omega^2_x x^2)$ allows
for an exact solution~\cite{connor68}.  The transmission and reflection
amplitudes are diagonal in the mode indices with
\begin{eqnarray}
t_{a} &=& e^{-i\alpha(\mathcal{E}_a)}\left\lbrace 1+e^{-2\pi\mathcal{E}_a}
\right\rbrace^{-1/2} \nonumber \\
r_{a} &=& -i\, e^{-i\alpha(\mathcal{E}_a)}e^{-\pi\mathcal{E}_a}\left\lbrace
1+e^{-2\pi\mathcal{E}_a} \right\rbrace^{-1/2}\, ,
\end{eqnarray}
where 
$\alpha(\mathcal{E})=\mathcal{E}+\arg\Gamma(\frac{1}{2}+i\mathcal{E})
-\mathcal{E} \ln |\mathcal{E}|$ in terms of the dimensionless energy
$\mathcal{E}_a=2[\varepsilon-\hbar\omega_y(a+\frac{1}{2})-V_0]/\hbar\omega_x$.
In the limit $\omega_y\gg\omega_x$, a good conductance quantization is achieved
\cite{buttiker90}. As expected for an arbitrary QPC, $|r^*t|$ is
appreciable only at the steps, and on the
$N$\textsuperscript{th} plateau 
\begin{equation}
\label{eq:deltag2-dho}
\delta g \simeq g^{(2)} \simeq -8 \pi^2\
 \sum_{a=0}^{N-1}\left|\mathcal{V}^{1,2}_{a,a}\right|^2\, . 
\end{equation}
Assuming a local tip with
$V(\mathbf{r})=V_\mathrm{T}\delta(\mathbf{r}-\mathbf{r}_0)$ and using the
semiclassical form of the scattering wavefunctions, we have
\begin{eqnarray}
\mathcal{V}^{1,2}_{a,a}&=&\frac{V_\mathrm{T}l|\phi_a(y_0)|^2
t^*_a}{2\pi\hbar^2\mathcal{K}/M}\times \nonumber \\
&& \hspace{-12mm}\left\lbrace r_a+\exp\left[-i\left(\frac{x_0}{l}\mathcal{K}
+2\mathcal{E}_a
\ln\left[\frac{\mathcal{K}+\frac{x_0}{2l}}{\sqrt{\mathcal{E}_a}}
\right]\right) \right]\right\rbrace \, ,
\label{eq:v12}
\end{eqnarray}
for the propagating modes, where $l=\sqrt{\hbar/2M\omega_x}$ and 
$\mathcal{K}=\sqrt{\mathcal{E}_a+(x_0/2l)^2}$. The $a$\textsuperscript{th} 
eigenstate $\phi_a(y)$ of a one-dimensional harmonic oscillator with frequency 
$\omega_y$ yields the lobes in the transverse direction that are ubiquitous in 
experiments and simulations. $\mathcal{V}^{1,2}$ has real and imaginary parts 
that oscillate as a function of $x_0$ with an overall decay proportional to 
$x_0^{-1}$ far from the constriction. The second-order correction
[Eq.~(\ref{eq:deltag2-dho})], dominant on the plateaus, is always negative, 
exhibits small fringes, and has an overall decay with $x_0^{-2}$, in agreement 
with Ref.~\cite{topinka01a}. 

\begin{figure}[tb]
\centerline{\includegraphics[width=\figwidth]{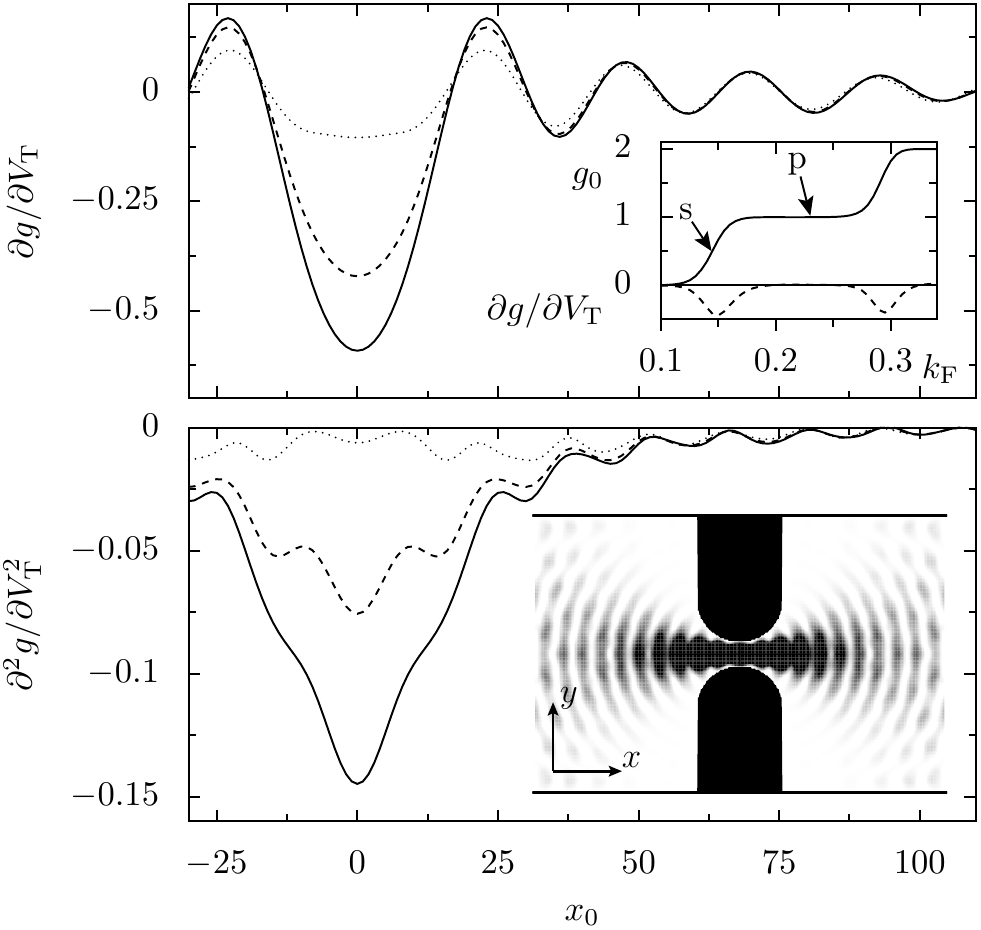}}
\caption{\label{fig:qpc} Top: $\partial g/\partial V_\mathrm{T}$ of the
conductance change with tip voltage for a hard wall QPC and a $\delta$-tip at
the first conductance step (point s of the conductance curve shown in the inset)
as a function of the tip position $x_0$.  The solid line is for 
$y_0=0$ (in the center); the dashed and dotted lines are for $y_0=4$ and 8,
respectively.  
Inset: The conductance as a function of the Fermi wave vector for the
unperturbed structure (solid line) and the linear correction (dashed) for
$x_0=0$, $y_0=4$.
Bottom: Similar, but for $\partial^2 g/\partial V_\mathrm{T}^2$ on the first
plateau (point p). Inset: Gray-scale plot of the same data as a function of $x_0$
and $y_0$ and the configuration with a minimal width $\Delta y = 20$
used in the numerical calculations. Anderson units are used
for $x_0$, $y_0$, $\Delta y$, $k_\mathrm{F}$, and $V_\mathrm{T}$.}
\end{figure}
Despite the simplicity of the harmonic saddle potential and its absence of mode
mixing, the results obtained are quite representative of a generic QPC's
behavior. This universality stems from the above mentioned transformation into
the transmission eigenmode basis, and was checked through numerical simulations
in a variety of confining potentials. The
results from recursive Green function algorithms~\cite{szafer89} for 
the double harmonic oscillator (not shown) reproduce our analytical 
calculations, while those corresponding to a
tight-binding model of a hard wall QPC are presented in Fig.~\ref{fig:qpc}.  The
change $\partial g/\partial V_\mathrm{T}$ is shown in the upper panel as a
function of the tip position  for a Fermi energy in the first conductance step
(point s), and as a function of $k_\mathrm{F}$ in the inset. Confirming our
analytical prediction, it is significant only in the step regions.  The lower
panel shows the change $\partial^2 g/\partial V_\mathrm{T}^2$ on the first
plateau (point p). In agreement with Eq.~\eqref{eq:deltag2}, it is a negative
correction that decreases with $x_0$ and $y_0$ further from the constriction
center. The inset shows the full dependence of $\partial^2 g/\partial
V_\mathrm{T}^2$ on the tip position. A checkerboard pattern consistent with
experimental findings \cite{jura09a} can be observed in the second-order
correction on the first plateau. The choice of a $\delta$ tip was made for
simplicity, but a smoother tip can be easily implemented. 

SGM measurements can be considerably influenced by disorder near the QPC. 
The conductance changes [Eqs.~(\ref{eq:deltag1}) and (\ref{eq:deltag2})] fully
incorporate the disorder through its effect on the scattering amplitudes and the
scattering wave-function basis for expressing $\mathcal{V}$.  The transformation
in mode space that singles out the propagating modes illustrates that the
conclusions regarding the relative importance of $g^{(1)}$ and $g^{(2)}$ remain
valid in the presence of disorder.  Since a disordered wire
containing a QPC can be mapped onto an unconstrained conductor with a renormalized
mean free path~\cite{beenakker94}, the random-matrix theory of quantum
transport~\cite{mello04} can be used to determine the statistical properties of
the SGM patterns far from the QPC.  The use of semiclassical 
methods~\cite{jalabert00} to evaluate the conductance corrections 
[Eqs.~(\ref{eq:deltag1}) and (\ref{eq:deltag2})] can yield an intuitive 
interpretation in terms of classical trajectories bridging the gap between
classical~\cite{topinka00a,topinka01a} and
quantum~\cite{pala08a,metalidis05,cresti06,he02} calculations and providing an 
understanding of the region near the QPC (where coherent branching and transient
behaviors occur). 

In summary, expressions for the lowest-order corrections to a nanostructure's 
conductance caused by a weak perturbation have been given.  Applied to the SGM
study of a QPC, the first-order corrections in the tip strength dominate at the
conductance steps and are suppressed on the plateaus.  The second-order corrections become dominant on the plateaus. The two expressions are of great interest since, even if most of the existing data have been obtained in the plateau regions, recent experiments~\cite{jura09a} also explored the behavior at the mode opening finding a different interference pattern.  It is also in this regime where the interaction effects~\cite{freyn08a,weinmann08a} are expected to show their signature.

The currently used tip potentials are strong enough to create a divot of 
depletion in the two-dimensional electron
gas~\cite{topinka00a,topinka01a,leroy05a,jura09a}, which is clearly beyond the
perturbative regime.  However, we anticipate that much of the interpretation and
many of the lessons derived from the perturbation calculations would still
apply.  In our quest for a rigorous interpretation of the SGM measurements, the
perturbative approach is an unavoidable landmark towards the understanding of
this fascinating problem.   Note that in experiments on nanostructures which do
not present conductance quantization, and for which therefore the linear
correction [Eq.~(\ref{eq:deltag1})] dominates the SGM response, the tip voltage
has been varied over large intervals~\cite{martins07a,pala08a} exhibiting the
expected linear behavior. 

\begin{acknowledgments}
We thank E.~J.~Heller, L.~Kaplan, J.-L.~Pichard, and C.~A.~Stafford for useful 
discussions and M.\ B\"uttiker for helpful correspondence. Financial support from the French National Research Agency ANR
(Project No.~ANR-08-BLAN-0030-02) and the U.S.~National Science Foundation
(PHY-0855337) is gratefully acknowledged.   We thank the Institute for
Nuclear Theory at the University of Washington for its hospitality and the DOE
for partial support at the initiation of this work. 
\end{acknowledgments}


\begin{thebibliography}{99}
\bibitem{fn-current-independent-of-x}
The $x$ position of $\mathcal{S}_x$ is irrelevant for the final result.
\bibitem{symmetry-of-dg}
The conductance change is invariant under a change of current direction as well
as under the changes $V(x,y)\to V(\pm x,\pm y)$ of the
perturbing potential, provided the unperturbed system verifies the corresponding
symmetries.
\bibitem{limit-low-tip-voltage}
The first-order correction Eq.~(\ref{eq:deltag1}) is not strictly zero on the 
plateaus.  For infinitesimal tip strength, it remains the dominant correction, 
but in measurements with lab-scale tip voltages, the second-order term becomes 
dominant.
\end{thebibliography}
\end{document}